\documentclass[10pt]{iopart}
\usepackage{bm,babel}
\usepackage{subfigure}
\usepackage{enumerate, pdflscape}
\usepackage{pgfplots,pgf,tikz} \usetikzlibrary{arrows}
\usepackage{booktabs, hyperref}
\expandafter\let\csname equation*\endcsname\relax
\expandafter\let\csname endequation*\endcsname\relax
\usepackage{amsmath, amssymb}
\usepackage{xcolor}
\definecolor{Gruen}{RGB}{64,136,39}
\definecolor{Gruen2}{RGB}{100,156,81}

\begin{document}
\title[A research-informed tool to visually approach Gauss' and Stokes' theorems]{A research-informed graphical tool to visually approach Gauss' and Stokes' theorems in vector calculus}

\author{L Hahn$^1$, S Blaue$^1$, P Klein$^1$}
\address{$^1$Physics Education Research, Faculty of Physics, University of G\"{o}ttingen, Friedrich-Hund-Platz 1, 37077 G\"{o}ttingen, Germany}

\ead{larissa.hahn@uni-goettingen.de}
\bibliographystyle{unsrt}
\begin{abstract}
Gauss' and Stokes' theorems are fundamental results in vector calculus and important tools in physics and engineering. When students are asked to describe the meaning of Gauss' divergence theorem, they often use statements like this: ``The sum of all sources of a vector field in a region gives the net flux out of the region". In order to raise this description to a mathematically sound level of understanding, we present an educational approach based on the visual interpretation of the vector differential operators, i.e. divergence and curl. As a starting point, we use simple vector field diagrams for a qualitative approach to connect both sides of the integral theorems, and present an interactive graphical tool to support this connection. The tool allows to visualise two-dimensional vector fields, to specify vector decomposition, to evaluate divergence and curl point wise, and to draw rectangles to determine surface and line integrals. From a meta-perspective, we situate this educational approach into learning with (multiple) representations. Based on prior research, the graphical tool addresses various learning difficulties of vector fields that are connected to divergence and curl. The tool was incorporated into the weekly lecture-based recitations of Physics II (electromagnetism) in 2022 and 2023, and we assessed various educational outcome measures. The students overall reported the tool to be intuitive and user-friendly (level of agreement $76\%$, $N=125$), considered it helpful for understanding and recommended its use for introductory physics courses (level of agreement $65\%$, $N=65$). 
\end{abstract}

\maketitle

\section{Introduction and educational background}

Vector fields are important mathematical objects used, for example, to describe electromagnetic fields or velocity fields of fluids. As such, they are fundamental in physics and other scientific disciplines. Vector fields are typically represented graphically by a set of arrows indicating the direction and magnitude of the field. For example, velocity vectors can be assigned to every particle of a moving liquid, representing the speed and the direction of the moving particle. The velocity of the liquid in each point $\vec{r}$ then defines a vector field $\vec{v}(\vec{r}\,)$. Likewise, vector fields can be described using mathematical equations. While equations are useful for quantitative calculations, vector field diagrams can present a lot of information at a glance, such as intensities or special field properties, for example, divergence or curl. These concepts play a major role in conservation laws that undergraduate physics students encounter in differential or integral form, for example, the continuity equation. In this context, a close connection to the integral theorem of Gauss is established which reflects the characteristics of a conservation law. Given a vector field $\vec{F}$ and a volume $V$ with a (closed) boundary surface $\partial V$ and its outer normal $\vec{n}$, Gauss' divergence theorem relates the volume integral of the divergence to the net flow across the boundary of the volume, 
\begin{equation}
    \int_V \textrm{div} \vec{F}\, dV = \oint_{\partial V} \vec{F} \cdot d\vec{n}\,\text{.}\label{eq:Gauss}
\end{equation}
As its name already suggests and Eq. \eqref{eq:Gauss} shows, the central concept of Gauss' theorem is the divergence; hence, it addresses several mathematical and physical concepts that are directly connected to divergence, for example, flux through boundaries or sources and sinks. In the context of conservation laws, another theorem is of great importance for physics applications, namely Stokes' theorem. For a vector field $\vec{F}$, a surface $A$ with outer normal $\vec{n}$, and a (closed) boundary $\partial A$ with vector path element $\vec{l}$, Stokes' theorem describes the relation between the surface integral of the curl to the circulation along the boundary of the area, 
\begin{equation}
     \int_A \textrm{rot} \vec{F}\cdot{} d\vec{n} = \oint_{\partial A} \vec{F} \cdot d \vec{l}\,\text{.}\label{eq:Stokes}
\end{equation}
In terms of form similar to Gauss' divergence theorem, Stokes' theorem focuses on the curl of a vector field, thus addressing corresponding concepts from mathematics and physics, i.e. partial derivatives and (surface and line) integration. Thus, both theorems are used to relate the local behaviour of the vector field to its global properties, providing valuable insights and facilitating calculations in various physical and mathematical contexts.\\
The importance of a sound understanding of vector calculus concepts is emphasised by a recent study which revealed that an extensive preparation in vector calculus is significantly related to students performance in an introductory electricity and magnetism course \cite{Burkholder2021}. This result is consistent with the findings of other studies which have shown that a solid mathematical understanding is essential for fully grasping the principles of physics \cite{smith2014, Singh&Maries}. In particular, Stokes' theorem fulfils all criteria of a threshold concept \cite{yusaf2017}, which, if fully understood by physics students, represents a fundamental milestone in their physics studies. Mathematics will become increasingly enriched with physical content as we improve our understanding of the corresponding mathematical concept. Without this understanding, a deeper comprehension of physics seems to be unattainable. In this article, we present a graphical vector field tool that takes empirical results on students' difficulties into account (Sect. \ref{sec.dificuties}) and supports a visual understanding of Gauss' and Stokes' theorems (Sects. \ref{sec.MRs}, \ref{sec.strategies}, and \ref{Sec.Stokes}). Following previous work in physics education, both sides of the theorems are addressed separately using multiple visual representations by renouncing a physics context as a first step towards a conceptual understanding (Sect. \ref{sec.simulation}). Finally, first results of students' evaluation on usability and perceived educational impact of the graphical tool
are reported (Sect. \ref{sec.tasks}). 

\subsection{Student difficulties with divergence, curl, and integral theorems} \label{sec.dificuties}
\begin{figure}[t!]
\centering
\begin{minipage}[h!]{0.4\textwidth}
\includegraphics[width=0.95\linewidth]{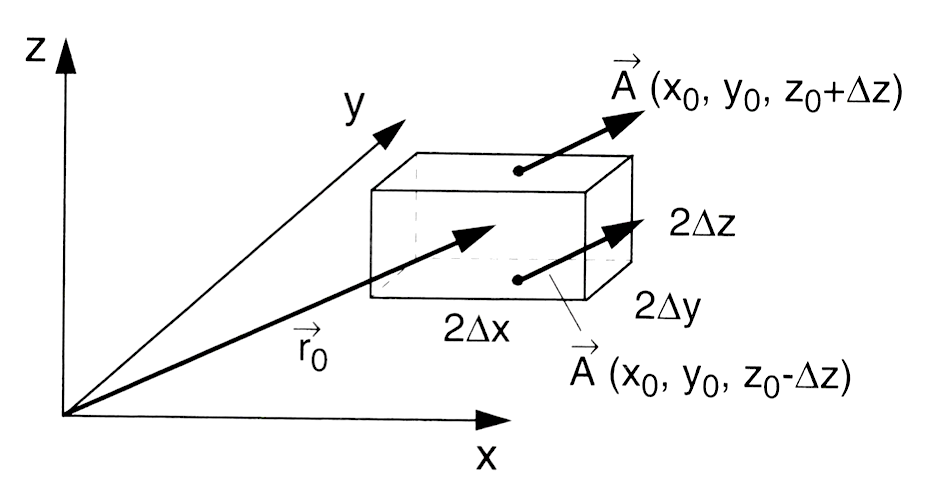}
\end{minipage}
\begin{minipage}[h!]{0.3\textwidth}
\includegraphics[width=0.95\linewidth]{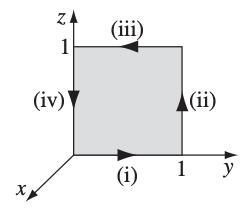}
\end{minipage}
\caption{Three-dimensional approach to Gauss' divergence theorem (left) and calculation-based approach to Stokes' theorem (right) in textbooks and lectures. \cite{Griffiths,Korsch}
    }
    \label{fig:3D}
\end{figure}
A comprehensive review by Smith (2014) found that introductory physics texts rarely provide conceptual details about the divergence theorem \cite{smith2014}; instead, textbooks often focus on mathematical treatments such as calculating both sides of Eqs. \eqref{eq:Gauss} and \eqref{eq:Stokes} explicitly or they use three-dimensional visualisations to proof the theorems (Fig. \ref{fig:3D}) \cite{Forster, Koenigsberger}. However, previous studies have shown that calculating divergence and curl hardly presents a problem for students; in contrast, the qualitative assessment is more challenging \cite{Singh&Maries, Klein2018, Bollen2015, Klein2019, Baily2015, Pepper, Bollen2018}. For example, students have difficulties to judge whether vector field diagrams have zero or non-zero divergence or curl, and they often do not understand that divergence and curl are local properties of the field. Various studies intensified this line of research and found different learning difficulties that are connected to the vector field representation, but transfer to differential vector operators and integral theorems, such as decomposing vector field arrows, interpreting partial vector derivatives, and differentiating field components and independent variables \cite{Klein2018, Bollen2015, Klein2019,Pepper, Bollen2017, Gire&Price, Hahn2023c, Klein2021}. In addition, Huffman \textit{et al.} (2020) found, that students are moderately able to literally interpret Gauss' theorem, but had great difficulty in formulating an intelligible and correct qualitative interpretation of the theorem, particularly of the left hand side integrand in Eq. \ref{eq:Gauss}. In particular, $48\%$ of the subjects stated that they were not able to formulate any qualitative interpretation \cite{Huffman2020}. Moreover, more than half of the physics students in a second year undergraduate course on vector calculus at the University of Bristol considered Stokes' theorem to be the most challenging topic of the course, because it was non-intuitive, difficult to visualise, lacked a physical analogy, and required an understanding of various other concepts \cite{yusaf2017}. In recent research, further, students were found to struggle with interpreting line, surface, and volume integrals and their differential elements, visualising an integration area, as well as balancing positive and negative fluxes through a test volume \cite{Klein2018, Pepper,Jones2020, Schermerhorn2019a, Schermerhorn2019b}.\\
Studies in the context of electromagnetism and electrostatics went beyond the investigation of mathematical issues with vector fields. These studies demonstrated that conceptual gaps regarding vector fields resulted in an improper comprehension of fundamental principles in physics, namely Maxwell’s equations \cite{Bollen2015, Bollen2016} and, in particular, Gauss' and Ampère's law \cite{LiSingh2017}. 

\subsection{Supporting students' visual understanding}\label{sec.MRs}
As pointed out above, an appropriate visual understanding of differential vector operators and integral theorems is considered to be crucial for approaching fundamental relationships in physics; hence, supporting students in this area turns out to be particularly important. In the aforementioned work, there is consensus that teaching this topic becomes more effective when first introducing it in a purely mathematics context \cite{Hernandez2023, Dray2023}, and by using explicit instructions and visual representations \cite{Bollen2018}. These suggestions relate to a general educational notion, that is, the promotion of students' representational competence \cite{Rau2017, Ainsworth2008, DeCock2012, Dufresne1997, Klein2017}. In this context, Klein \textit{et al.} (2018) introduced explicit text-based instructions for visually interpreting the divergence of vector field diagrams \cite{Klein2018}. They presented strategies that refer to both sides of Eq. \eqref{eq:Gauss} and emphasise different mathematical concepts of divergence, namely partial derivatives and flux through boundaries \cite{Klein2018}. Both strategies are revisited in Sect. \ref{sec.strategies}. Even though students appreciated the visual approach towards divergence and improved their conceptual understanding, many students indicated that ``additional visual aids, such as sketches of the vector decomposition [...] would have improved [their] performance" \cite[p. 010116-14]{Klein2018}. In three experimental follow up-studies, it was shown that adding visual cues and drawing activities to the instruction in fact resulted in better learning outcomes \cite{Klein2019, Klein2021, Hahn2023}.\\ 
In addition, Bollen \textit{et al.} (2016) advocate modern instructional approaches---including technology---when teaching vector calculus, since traditional instruction, i.e. text-based instruction, is insufficient to enable a complete understanding of differential operators \cite{Bollen2016}. This claim is in line with Wieman's statement about the implementation of modern technology to foster effective science education \cite{Wieman2007}. Following this direction, some researchers and lecturers developed interactive and dynamic visualisation tools and simulations to support a conceptual understanding of vector fields and vector calculus. Such tools allow to visualise two- and three-dimensional vector field diagrams, calculate, plot, and indicate divergence and curl for a given field \cite{Budi2018, Ozgun2015, WaterlooSim}, and explore vector line integrals by visualising a curve in a given field and calculating circulation and flux of the field around the curve \cite{Campuzano2019}. However, there is no tool integrating all of those field concepts reported in educational literature. Here, we present such a graphical vector field tool that illustrates both sides of the Gauss' and Stokes' theorems using two-dimensional vector field plots. In contrast to the aforementioned work, this tool explicitly takes previous research on students difficulties with vector fields into account and builds upon a visual approach to Gauss' and Stokes' theorems that has been proved to support students' learning about vector fields and differential vector operators \cite{Klein2018, Klein2019, Klein2021, Ruf2022}.

\section{Towards a visual interpretation of divergence and Gauss' theorem}\label{sec.strategies}

Below, we explain a visual approach to Gauss' and Stokes' theorems using simple examples and highlight the underlying educational considerations. We refer to the graphical vector field tool \cite{Webapplication}, so that interested readers can follow our reasoning. All features of the tool will be summarised in Sect. \ref{sec.simulation} and the references here relate to the overview in Tab. \ref{tab:features}. 
\newline

\noindent Gauss' (divergence) theorem combines two intuitions of divergence, that is, the local interpretation of sources or sinks at specific points, given on the left-hand side of Eq. \eqref{eq:Gauss}, and the global interpretation as the net outward flow of an area which is represented on the right-hand side of Eq. \eqref{eq:Gauss}. Both interpretations can be applied to a vector field representation for a qualitative evaluation of divergence \cite{Klein2018, Klein2019, Klein2021}. Hereafter, we demonstrate both interpretations by applying the following educational simplifications:
\begin{enumerate}
\item We restrict our treatment to two dimensions. Thus, the test volume becomes a test area with a boundary in one dimension, whose edges can be varied arbitrarily. By doing so, we avoid the difficulty of representing three-dimensional test volumes and bounding surfaces.
\item We use the Cartesian coordinate system with the Cartesian coordinates $x$ and $y$ and the unit vectors $\hat{e}_x$ and $\hat{e}_y$; the field components are functions of $x$ and $y$. 
\item The field components depend only linearly on the coordinates $x$ and $y$. 
\item We rely on a qualitative approach, so the absolute values of the divergence and the surface integrals are not focused on.
\item We further restrict our treatment to vector fields with constant divergence (either zero or non-zero). 
\end{enumerate}
Concerning (iii) and (v), please note that these simplifications can be removed during the learning process. The graphical tool also represents vector fields with complex dependent components and non-constant divergence (Sects. \ref{sec.simulation} and \ref{sec.tasks}). \\
 
\subsection{An example with zero divergence}
\begin{figure}[t!]
    \centering
    \includegraphics[width=\linewidth]{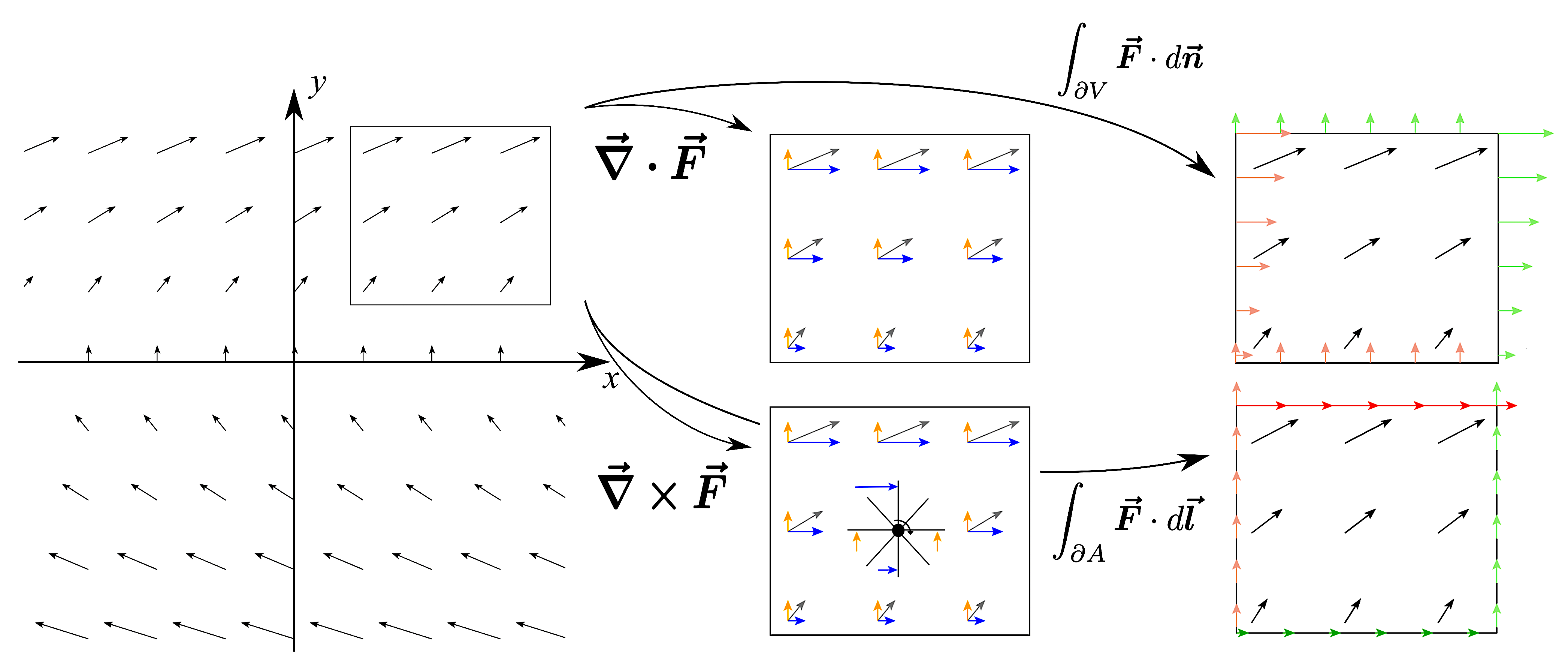}
    \caption{Graphical representation of the two-dimensional vector field $\vec{F}_1$ (left). A region is highlighted to evaluate divergence $\vec{\nabla}\cdot\vec{F}$ and curl $\vec{\nabla}\times\vec{F}$ locally using partial derivatives (middle) and a paddle wheel as well as globally by the flow through the boundary (top right) and by the circulation around the boundary (bottom right). In the middle, $x$ components $F_{1,x}$ are presented with blue arrows and $y$ components $F_{1,y}$ are in orange. On the right, an outward flow (top) and a (mathematically) positive circulation are visualised in green, and an inward flow as well as a (mathematically) negative circulation are presented with red arrows. 
    }
    \label{fig:Example1}
\end{figure}

Fig. \ref{fig:Example1} (left) shows the vector field diagram of $\vec{F}_1$ defined as 
\begin{equation*}
    \vec{F}_1(x,y) = \left( \begin{array}{c}
         F_{1,x}(x,y) \\
          F_{1,y}(x,y)
    \end{array}\right)= \left( \begin{array}{c}
         y \\
         1 
    \end{array}\right)\,\text{.}
\end{equation*}
Using the definition of divergence in Cartesian coordinates,
\begin{equation}
   \textrm{div}\vec{F}=\vec{\nabla}\cdot{}\vec{F}=\frac{\partial F_x}{\partial x}+\frac{\partial F_y}{\partial y}={\partial_x}{F_x}+{\partial_y}{F_y}\,\text{,}
   \label{eq:div}
\end{equation}
the left-hand side of Eq. \eqref{eq:Gauss} can be interpreted visually by tracing the changes in the field components. Vividly speaking, we need to visually examine the partial derivatives by evaluating the changes of $F_x$ (the $x$ component of the vectors) in $x$ direction, and likewise  the changes of $F_y$ in $y$ direction (Fig. \ref{fig:Example1}, top middle; blue arrow $F_x$, orange arrows $F_y$). For $\vec{F}_1$, we find ${\partial_x}{F_{1,x}}=0$ and ${\partial_y}{F_{1,y}}=0$, hence $ \textrm{div}\vec{F}_1=0$. This procedure leads to the same result for each point $\left(x,y\right)$ of the field $\vec{F}_1$. In general, the local evaluation of divergence indicates local sources and sinks of the vector field. Therefore, the surface integral of the (zero or non-zero) divergence can be referred to as the sum of all sources and sinks within a considered area.\\
Regarding the right-hand side of Eq. \eqref{eq:Gauss}, divergence can also be evaluated by balancing the inward and outward flow through the boundaries of the test area. Using rectangles as test areas, Fig. \ref{fig:Example1} (top right) shows the projection of $\vec{F}_1$ onto the (outer) normal $\vec{n}$. Referring to the surface integral as the sum of all scalar products $\vec{F}\cdot d\vec{n}$ at each point of the boundary, one can qualitatively evaluate the net flux through the test area. Concerning $\vec{F}_1$, it then can be observed that for every arrow entering the test area, there is an equivalent arrow leaving it (for symmetry reasons). Hence, we can conclude that $\int_{\partial V} \vec{F}_1 \cdot d\vec{n}=0$. Similar to the local interpretation, this procedure leads to the same result for any chosen area. The described properties and concepts can be traced with the graphical tool (Tab. \ref{tab:features}, \textbf{5} and \textbf{6}). 
\subsection{An example with non-zero divergence}
\begin{figure}[t!]
    \centering
    \includegraphics[width=0.6\linewidth]{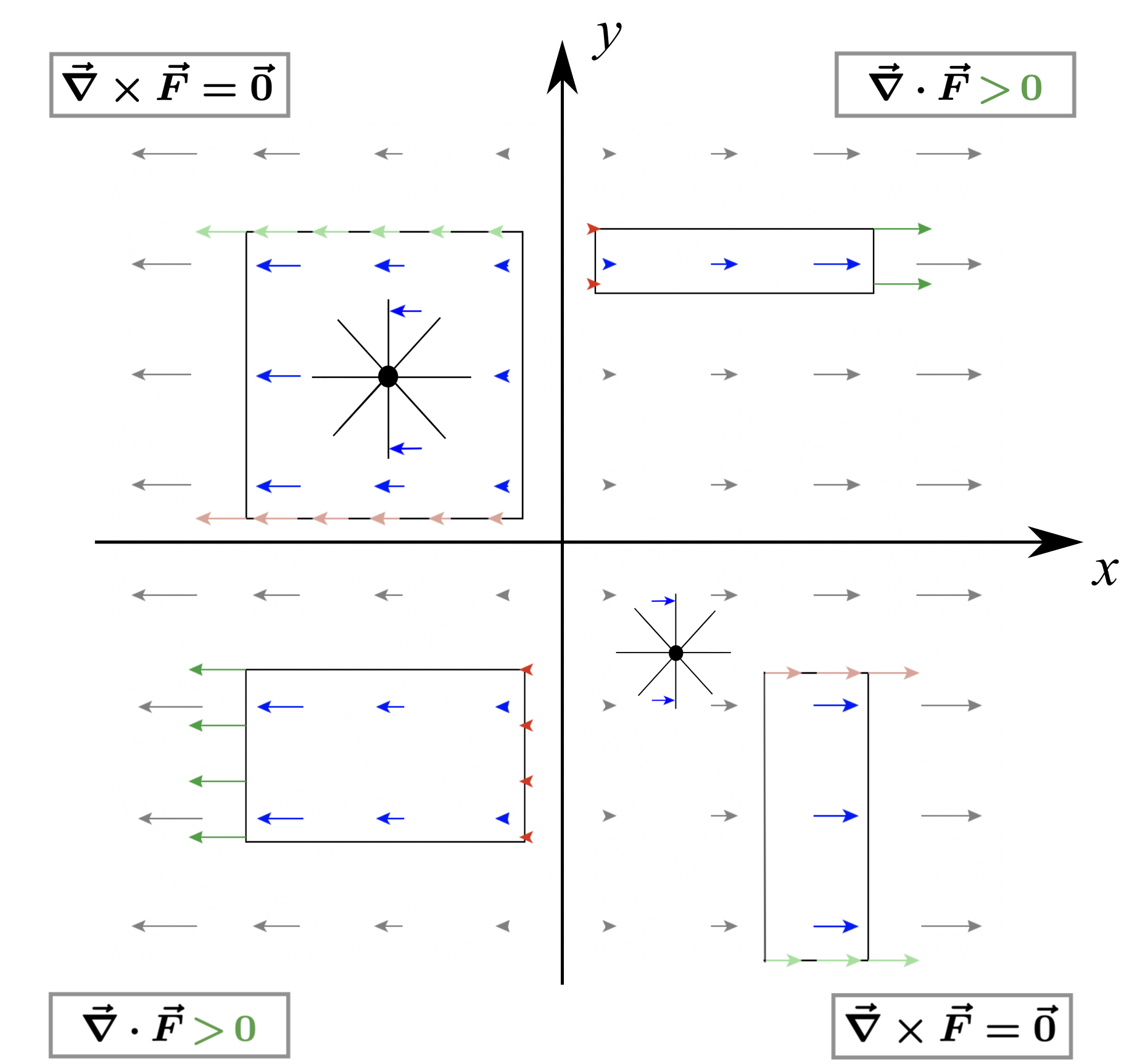}
    \caption{Graphical representation of $\vec{F}_2$ with $\vec{F}_{2,y}=0$. In the test areas on the top right and the bottom left of the field (quadrant I and III) divergence is evaluated by tracing changes in the $x$ component $\vec{F}_{2,x}$ in $x$ direction, i.e., along a row (blue arrows) or by balancing the flow through the boundaries (green and red arrows). In the test areas on the top left and the bottom right of the field (quadrant II and IV) the fields' curl is determined by  evaluating changes in the $y$ component $\vec{F}_{2,x}$ in $y$ direction, i.e., along a column (blue arrows), by balancing circulation around boundaries (green and red arrows), and by inserting a paddle wheel.
    }
    \label{fig:Example2}
\end{figure}

For another example, we apply the two visual strategies to $\vec{F}_2$, which is defined as
\begin{equation*}
    \vec{F}_2(x,y) = \left( \begin{array}{c}
         F_{2,x}(x,y) \\
          F_{2,y}(x,y)
    \end{array}\right) = \left( \begin{array}{c}
         x \\
         0 
    \end{array}\right)\,\text{.}
\end{equation*} 
Because it is $F_{2,y}=0$, there is no change of $F_{2,y}$ in $y$ direction (Fig. \ref{fig:Example2}, quadrant I and III), hence ${\partial_y}{F_{2,y}}=0$. Tracing $F_{2,x}$ in positive $x$ direction (Fig. \ref{fig:Example2}, quadrant I and III, blue arrows), one observes an elongation, i.e. ${\partial_x}{F_{2,x}}>0$. Thus, the qualitative evaluation via partial derivatives (Eq. (\ref{eq:div})) yields a positive divergence. For the global approach regarding the net flow through a test area, we first consider the horizontal boundaries. Since $F_{2,y}$ is zero, here, all field components are perpendicular to the outer normal vector $\vec{n}$, so the net flux through the horizontal boundaries equals zero (Fig. \ref{fig:Example2}, quadrant I and III). For the vertical boundaries, the outer normal vectors are parallel to the field vectors, and we observe a positive outward flow; therefore, the integral is non-zero and positive. In Fig. \ref{fig:Example2}, the visual interpretation is applied to different quadrants of the field yielding $\vec{\nabla}\cdot\vec{F}_2>0$ for every spot in the field.

\subsection{Visual connection of both sides of the equation}\label{sec.connection}
Let us reconsider both examples to establish an explicit connection between both sides of Eq. \eqref{eq:Gauss}. Regarding $\vec{F}_1$, since $F_{1,x}$ does not change in $x$ direction (Fig. \ref{fig:Example1} blue arrows, top middle), the flow through the left and the right vertical boundaries is equal and thus cancels out, so the net flow through the vertical boundaries equals zero (Fig. \ref{fig:Example1} top right). The same occurs for $F_{1,y}$, where we observe no change of $F_{1,y}$ in $y$ direction (Fig. \ref{fig:Example1} orange arrows, top middle) and, thus, the lower and upper horizontal boundaries register no net flow (Fig. \ref{fig:Example1} top right). In total, this results in a net flow that equals zero out of the area. In other words, to register a total net flow comparing two vertical (horizontal) boundaries, there must be a change of the $x$ component in the $x$ direction (of the $y$ component in the $y$ direction). This is evident in the second example, $\vec{F}_2$, where we observe an elongation of $F_{2,x}$ in the $x$ direction that results in a positive flow outward from the test area. In summary, we have qualitatively evaluated the divergence of the vector field $\vec{F}_1$ as zero and of the vector field $\vec{F}_2$ as positive, whereas both intuitions of divergence can be connected by Gauss' divergence theorem. On closer inspection, however, it becomes clear from Eq. \eqref{eq:div} that the divergence can be zero, even though a change of the vector field components in $x$ or $y$ direction exists. This is the case when the partial derivatives cancel out exactly. For example, for a vector field $\vec{F}_3=x\hat{e}_x-y\hat{e}_y$, it is  $\textrm{div}\vec{F}_3=0$, although the partial derivatives are non-zero, i.e., $\partial_x F_{3,x}=1$ and $\partial_y F_{3,y}=-1$. Using the graphical tool, we observe, that the positive flow through the left edge of a rectangle is compensated with a negative flow through the top or bottom edge, not through the opposite, right edge. Beyond $\vec{F}_3$, a positive inward flow through one edge can also flow out through all other edges. And vice versa: a negative outward flow through one edge can be compensated by the inflow through all other edges. For this reason, it is important to consider and balance the flux over the entire test area.   

\section{Visually interpreting curl and Stokes' theorem}\label{Sec.Stokes}
Comparable to Gauss' divergence theorem, Stokes' theorem combines two intuitions of curl, that is, the local interpretation of vortexes at specific points (left-hand side of Eq. \eqref{eq:Stokes}), and the global interpretation of the circulation around a closed surface (right-hand side of Eq. \eqref{eq:Gauss}). Basically, by administering the aforementioned simplifications, by transferring the visual interpretation of the partial derivatives from divergence to curl and instead of balancing the projection on the outer normal but on the vector path element, the proposed visual approach to Gauss' theorem can be applied to Stokes' theorem. However, since curl is represented by a vector with a direction perpendicular to the vector field plane, we formally refer to vector fields $\vec{F}(x,y,z) = (F_x, F_y, 0)$ in three dimensions, whereas the $z$ component $F_z$ equals zero. Then, by referring to the definition of curl in Cartesian coordinates in two dimensions,
\begin{equation}
   \textrm{rot}\vec{F}=\vec{\nabla}\times\vec{F}=\left(\frac{\partial F_y}{\partial x}-\frac{\partial F_x}{\partial y}\right)\hat{e}_z\,\text{,}
   \label{eq:rot}
\end{equation}
the visual interpretation of the partial derivatives through tracing the changes in the field components can be applied for an interpretation of the left-hand side of Eq. \eqref{eq:Stokes}. Compared to the visual interpretation of divergence, the main change here is that the ``mixed" changes of the $F_x$ component in $y$ direction and the $F_y$ component in $x$ direction are examined (Eq. (\ref{eq:rot})). For $\vec{F}_1$, we find ${\partial_x}{F_{1,y}}=0$ and ${\partial_y}{F_{1,x}}=1$, hence $\textrm{rot}\vec{F}_1=-1\hat{e}_z$ (Fig. \ref{fig:Example1}, bottom middle). Further, this approach yields $\textrm{rot}\vec{F}_2=\vec{0}$ for $\vec{F}_2$ (Fig. \ref{fig:Example2}, quadrant II and IV, blue arrows and paddle wheel). This procedure leads to the same result for each point $\left(x,y\right)$ of the field $\vec{F}_1$ and $\vec{F}_2$, respectively. Analogously to the interpretation of divergence, the local evaluation of curl indicates local vortexes of the vector field. For visualisation purposes, the interaction of the field components with an inserted paddle wheel (Fig. \ref{fig:Example1}, bottom middle) can be considered (Tab. \ref{tab:features}, \textbf{9}) \cite{Bollen2018,Jung2012, Huang2013}. Then, changes of the field components lead to a rotation of the paddle wheel, because the field's interaction with the wheel is no longer balanced for all paddles of the wheel. As the wheel rotates around a rotation axis perpendicular to the $x$-$y$-level of the field, curl is associated with a vector quantity. Further, by referring to the left-hand side of Eq. \eqref{eq:Stokes} all local vortexes of the field are summed up within the considered area through the surface integral. When using rectangles in the $x$-$y$-level as test areas, the curl vector $\textrm{rot}\vec{F}$ and the areas outer normal vector $\vec{n}$ are parallel oriented in $z$ direction. Therefore, $\textrm{rot}\vec{F}\cdot d\vec{n}$ is qualitatively interpreted as the value of $\textrm{rot}\vec{F}$. \\
Regarding the right-hand side of Eq. \eqref{eq:Stokes}, the curl can also be evaluated by balancing the circulation around the boundary of the test area. Here, a positive circulation is referred to as being mathematically positive, i.e. counterclockwise, directed and a negative circulation indicates a mathematically negative, clockwise direction. By referring to the line integral as the sum of all scalar products $\vec{F}\cdot d\vec{l}$, i.e. the projection of the vector field onto the vector path elements $d\vec{l}$, at each point of the boundary, the net circulation around the test surface be can evaluated qualitatively (Fig. \ref{fig:Example1} and \ref{fig:Example2}). Taking the approaches of partial derivatives and circulation into account, we can establish a connection between both sides of Eq. \eqref{eq:Stokes}. For $\vec{F}_1$, it can be observed that the change of the $x$ component $\vec{F}_{1,x}$ in $y$ direction (Fig. \ref{fig:Example1} blue arrows, bottom middle) leads to an imbalance of positive and negative circulation along the top and bottom edge of the test area (Fig. \ref{fig:Example1} bottom right; Tab. \ref{tab:features}, \textbf{5} and \textbf{7}). Hence, we find a negative net circulation around the surface (Fig. \ref{fig:Example1} bottom right). By referring to the paddle wheel approach, one observes a clockwise (mathematically negative) rotation of the wheel due to the imbalanced impact of the field on the top and the bottom paddle (Fig. \ref{fig:Example1} bottom middle). From that, we can again conclude that the net circulation is negative. For $\vec{F}_2$, positive and negative circulation are balanced for all edges of the rectangle, thus we observe a zero net circulation. Similar to the local interpretation, this procedure leads to the same result for any chosen area. As for Gauss' divergence theorem, a positive circulation along one edge of the rectangle does not necessarily have to be compensated with a negative circulation along the opposite edge, i.e. the curl can be zero, even though vector field components change in $x$ or $y$ direction. Therefore, balancing all contributions of the circulation around the whole boundary is of particular importance (Sect. \ref{sec.connection}).

\section{Features of the graphical vector field tool}\label{sec.simulation}

\begin{figure}[t!]
\centering
\includegraphics[width=\linewidth]{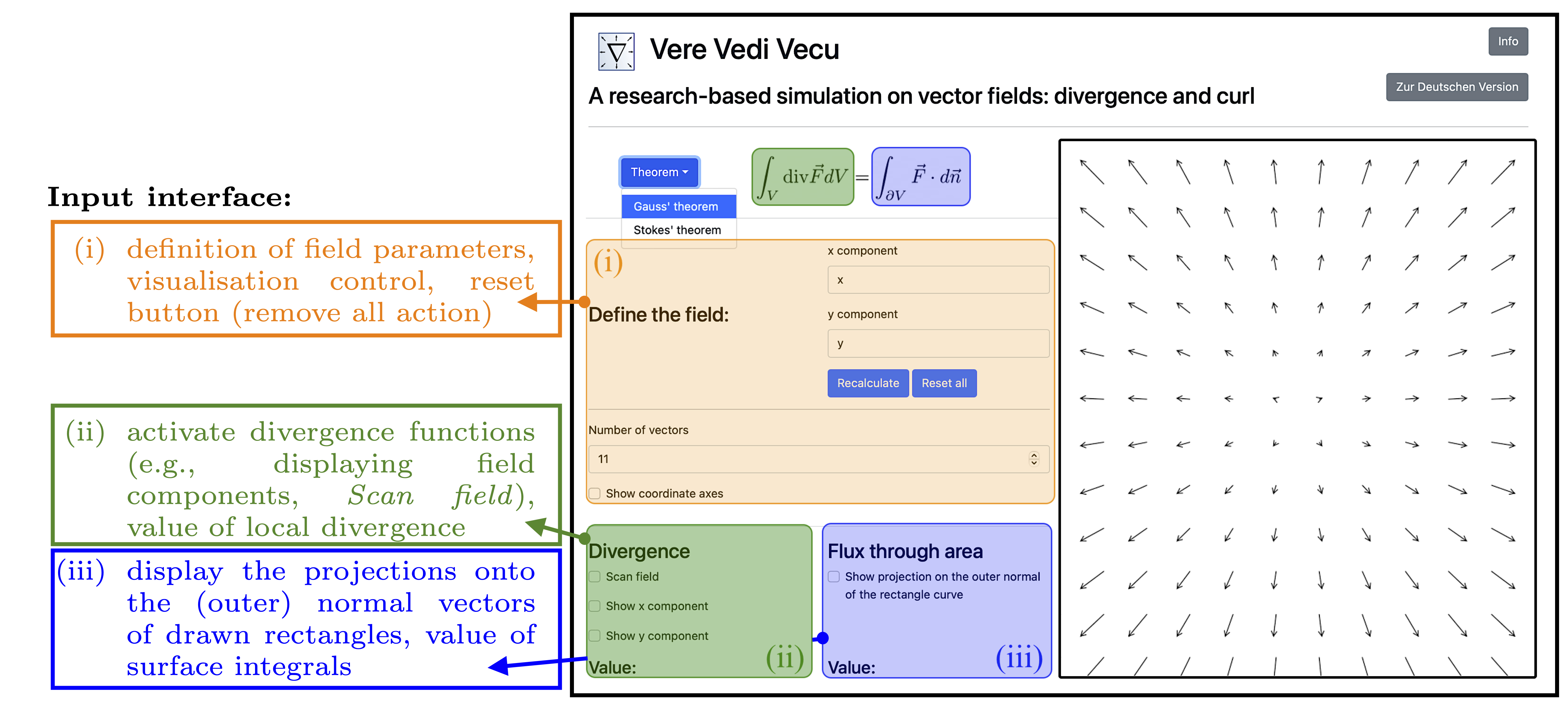}
\caption{User interface of the graphical tool with Gauss' divergence theorem selected. The features of the input interface (left) are explained left beside. The vector field plot is shown on the right and allows various user actions (e.g., drawing rectangles and drawing vector path). If Stokes' theorem is selected, areas (ii) and (iii) analogously display curl and circulation along a curve, with the only change that an additional check box in (iii) allows to insert a paddle wheel.
    }
    \label{fig:oberflache}
\end{figure}

In light of previous research on students' difficulties about vector fields and divergence, we developed a graphical tool that takes these findings into account by referring to the approaches explained in Sects. \ref{sec.strategies} and \ref{Sec.Stokes}. The features and how they relate to student difficulties are summarised in Table \ref{tab:features}. A German and an English version of the tool for desktop computers and laptops is freely available online \cite{Webapplication}. Fig. \ref{fig:oberflache} illustrates the user interface of the graphical tool with the key functions being highlighted. A short introductory video presenting the visualisation format and all features of the tool related to the visual interpretation of the field concepts is available online \cite{Video1}. 

\begin{table}[b!]
\centering
\caption{Integration of educational considerations in the graphical tool. The literature refers to student difficulties with properties of (two-dimensional) vector fields.}
\footnotesize 
\label{tab:features}
\begin{tabular}{p{6cm} p{6cm}}
\hline \hline 
 Physics property or concept of vector fields$^{\textrm{a}}$ & Feature of the graphical tool\\ \hline \addlinespace[3pt]
\textbf{1} A selection of arrows represent the vector field. They start at equidistant grid points and are not centred on their location. \cite{Bollen2017, Gire&Price}& Visual representation of the field with zoom function; Only few representative vectors are shown. \\ \addlinespace[3pt]
\textbf{2} A vector field is dense in space. Between any two arrows, there exists a field vector rather than space being empty. \cite{Bollen2017, Gire&Price} & Right click on any spot reveals the arrow starting at that spot. When the box \textit{Scan field} is activated, the smooth continuation of the field is experienced when inserting additional arrows (vector path) by clicking, holding, and moving the mouse. \\\addlinespace[3pt]
\textbf{3} A vector field has components $F_x$ and $F_y$; Both components can depend on input variables ($x$ and $y$). \cite{Bollen2017,Gire&Price,Hahn2023c,Barniol2014,  Knight1995} & Clear separation of $F_x$ and $F_y$ in the control panel; User must enter $x$ and $y$ explicitly; Decomposition of vectors can be shown. \\ \addlinespace[3pt]
\textbf{4} Vector decomposition. \cite{Klein2018, Bollen2017,Hahn2023c,Barniol2014,  Knight1995} & Highlighting $F_x$ and $F_y$ within a user-defined rectangle and along a vector path using different colours.\\\addlinespace[3pt]
\textbf{5} Partial vector derivatives. \cite{Klein2018, Pepper} & Highlighting the direction of change within a user-defined rectangle and along a vector path; Inserting a paddle wheel.\\\addlinespace[1pt]
\textbf{6} Surface integral/Flux through boundary (sum over scalar products between vector field and outer surface normal). \cite{Klein2018, Pepper} & Highlighting the projection on the outer normal vector; Colour coding for positive and negative flux.\\\addlinespace[3pt]
\textbf{7} Vector line integral/circulation around boundary (sum over scalar products between vector field and vector path element). \cite{Pepper,Jones2020} & Highlighting the projection on the vector path element; Colour coding for positive and negative circulation.\\\addlinespace[3pt]
\textbf{8} Divergence is a scalar quantity. \cite{Bollen2015, Baily2015, Bollen2018} & Calculating numerical values. \\\addlinespace[3pt]
\textbf{9} Curl is a vector quantity and has a direction. \cite{Bollen2015, Baily2015, Bollen2018} & Spinning of the paddle wheel around a rotation axis. \\\addlinespace[3pt]
\textbf{10} Divergence and curl are different for every spot in the field. \cite{Baily2015, Bollen2018} & Tracing vector paths (\textit{Scan field}); Calculating local divergence and curl; Moving test area; Moving paddle wheel (curl).\\\addlinespace[3pt]
 \hline \hline 
\end{tabular}
\end{table}

\subsection{Vector field representation}
\noindent  To define a vector field, the user can enter both field components separately using Cartesian coordinates $x$, $y$, scalars, and mathematical operations. The interface makes a clear distinction between the field components ($F_x$ and $F_y$) and the input variables ($x$ and $y$; Tab. \ref{tab:features}, \textbf{3}). For any vector field, only a representative selection of arrows is displayed, ranging from 5 up to 25 vectors in each line or column, and each vector starts at equidistant grid points (Tab. \ref{tab:features}, \textbf{1}). For a better overview, very short vectors are completely blanked out. Changing the slider affects the number of vectors as you zoom in and out of the diagram to strengthen the understanding of the diagram as a collection of representatives (Tab. \ref{tab:features}, \textbf{1}). When the box \textit{Scan field} is activated, clicking with the mouse button in the empty area between any two vectors shows a vector path of the invisible arrows, thus indicating a smooth continuation of the field (Tab. \ref{tab:features}, \textbf{2}). In addition, holding the mouse button and moving the cursor across the field displays the field arrow of the cursor track, once more emphasising the fields' smooth continuation (Tab. \ref{tab:features}, \textbf{2}). To facilitate the connection between the algebraic equation and the vector field diagram, the Cartesian coordinate axes can be faded in. For visualisation purposes, the axis scale does not correspond to the displayed vector length. However, since our approach is qualitative in nature, exact numerical values are not necessary to visually understand the divergence theorem, and we recommend to hide the axes.

\subsection{Features to visually understand partial vector derivatives, flux through boundaries, and circulation around boundaries}
\begin{figure}[t!]
\centering
\includegraphics[width=\linewidth]{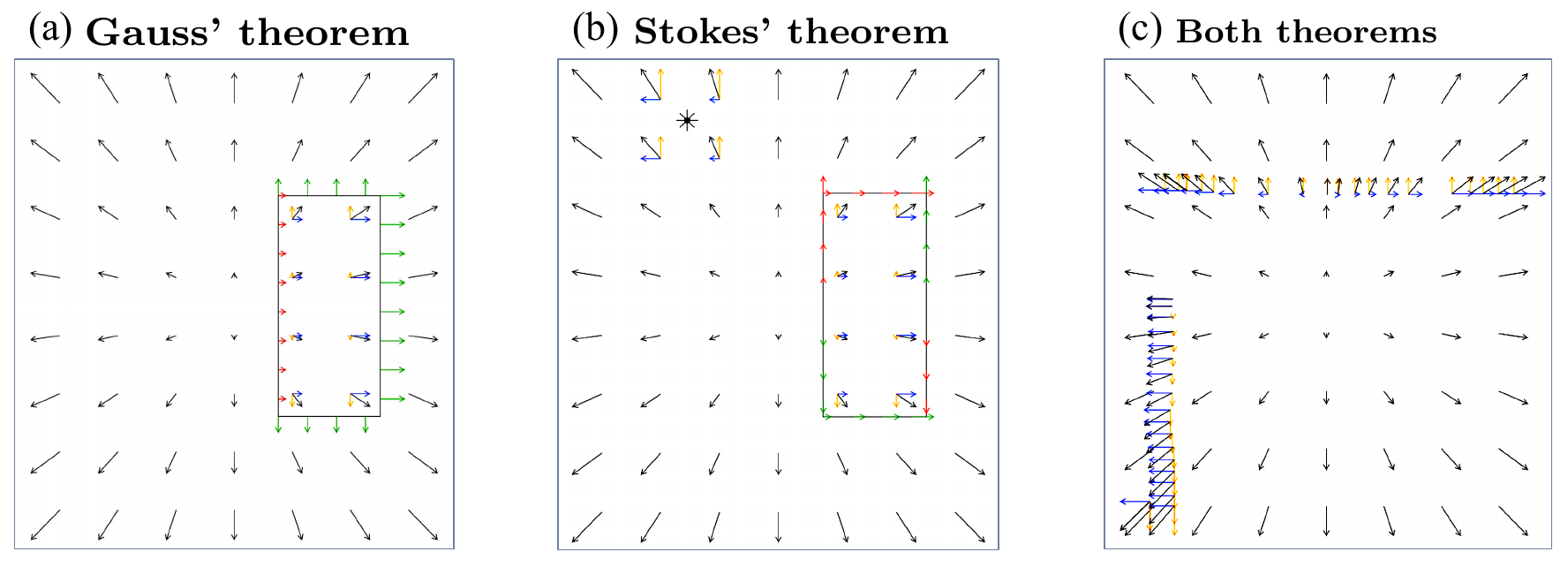}
\caption{Overview about the main features of the graphical tool for (a) Gauss' divergence theorem setting, (b) Stokes' theorem setting, and (c) both settings. (a) Displaying of vector decomposition (blue and orange arrows) and flux trough boundaries (green and red arrows) for drawn rectangles; (b) Displaying of vector decomposition (blue and orange arrows) and circulation around boundaries (green and red arrows) for drawn rectangles, as well as inserting a paddle wheel (top) with shown vector decomposition around its position; (c) Tracing arrows and its components in the field (\textit{Scan field} function).
    }
    \label{fig:features}
\end{figure}

Besides illustrating some basic properties of vector field diagrams, the key feature of the graphical tool consists of approaching the different visual interpretations associated with Gauss' and Stokes' theorems. At the top of the tool, the user can chose the theorem to be studied. Then, the bottom part of the tool adjusts either showing check boxes and values for divergence and flux through area or curl and circulation along a curve.\\
Independent of the integral theorem, the cursor can be used to draw a rectangle of any size and at any location within the vector field diagram, which serves as the test area for evaluating divergence and curl. For all vectors within that area, the vector decomposition for both components can be highlighted separately for $F_x$ and $F_y$ in different colours (Tab. \ref{tab:features}, \textbf{4}; Fig. \ref{fig:features}(a) and (b)). Since previous research has reported that many students struggle with vector addition and decomposition \cite{Klein2018, Bollen2017,Barniol2014,  Knight1995}, the auxiliary arrows help overcome these problems and also assist in evaluating partial derivatives and, thus, divergence and curl with respect to the left-hand side of Eqs.~\eqref{eq:Gauss} and ~\eqref{eq:Stokes} (Tab. \ref{tab:features}, \textbf{5}). Further, when the box \textit{Scan field} is activated, the user can click the mouse button and move the cursor across the field. Then, a vector track highlighting the vector decomposition is displayed, which enables the manual evaluation of changes in the vector components, hence checking partial vector derivatives (Tab. \ref{tab:features}, \textbf{5}; Fig. \ref{fig:features}(c)). For curl, inserting a paddle wheel also supports the identification of changes in the vector components (Fig. \ref{fig:features}(b)).\\
For drawn rectangles, beyond displaying vector decomposition, the projection of the vector field onto the (outer) normal vector or on the vector path element can be visualised by activating the corresponding check boxes, in Gauss' or Stokes' theorems setting respectively. For Gauss' theorem, the inward (negative) flow is represented by red arrows and the outward (positive) flow is represented by green arrows (Tab. \ref{tab:features}, \textbf{6}; Fig. \ref{fig:features}(a)). For Stokes' theorem, red arrows indicate a clockwise (mathematically negative) circulation and green arrows indicate an anticlockwise (mathematically positive) circulation (Tab. \ref{tab:features}, \textbf{7}; Fig. \ref{fig:features}(b)). Furthermore, the values of the surface and the line integrals as well as divergence and curl are shown numerically (Tab. \ref{tab:features}, \textbf{8} and \textbf{10}). Thus, the qualitative connection between both sides of Gauss' and Stokes' theorems can be established using the aforementioned strategies.\\
The user can also move one side of a fixed test area, enlarging or reducing its size. The tool then dynamically adjusts the integral values and the visual highlights. When the user changes the number of arrows, the tool keeps the coordinate range and the drawn test area remains at a fixed position. Consequently, the flux through the boundaries or the circulation around the boundaries does not change when the slider is used. This demonstrates that a higher number of arrows does not affect the fields' properties as the vector field and the considered test area remain the same.
\subsection{Features to understand local dependence of divergence and curl}
Following the study of Baily \textit{et al.} (2015), students often have difficulties to understand that divergence and curl can be different for each spot in the field \cite{Baily2015}. To visualise this aspect, the graphical tool allows to move a drawn test area across the field (Tab. \ref{tab:features}, \textbf{10}). By doing so, the vector decomposition, the in- and outward flow, the circulation, and the value of the surface and the line integral adjust dynamically. To further address the difficulty of local dependence of divergence and curl a left mouse click at any point within the field displays divergence or curl ($\textrm{rot}\vec{F}\cdot{}\hat{e}_z$) at this spot numerically (\textit{Scan field} deactivated). As described above, the box \textit{Scan field} can be activated to trace a vector path drawn by holding the left mouse button (Fig. \ref{fig:features}(c)). Then, the values of divergence and curl continuously adjust. Every path remains until the box \textit{Scan field} is deactivated or all actions are reset. Thus, the user can scan the field, for example, to identify characteristic points (e.g., maximum or zero divergence or curl). In addition, for Stokes' theorem, a paddle wheel can be inserted and moved within the field. Its rotation indicates whether the curl is zero or non-zero. For non-zero curl, the curls' sign (positive or negative) is indicated by the rotational direction. The rotation of the wheel around a rotation axis perpendicular to the field, further, refers to curl as a vector quantity, a common student difficulty (Tab. \ref{tab:features}, \textbf{9}) \cite{Bollen2015, Baily2015, Bollen2018}.

\section{Implementation in university courses and evaluation}\label{sec.tasks}

\begin{figure}[t!]
\centering
\includegraphics[width=0.95\linewidth]{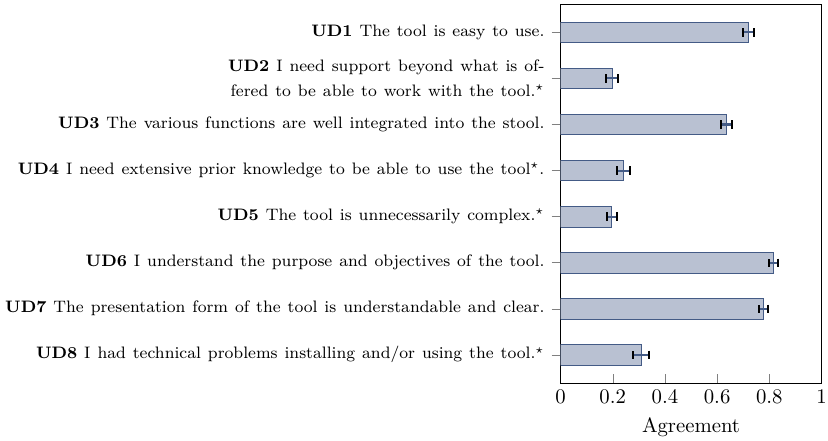}
\caption{Evaluation results of the usability and design scale for $N=125$ students (Cronbach's alpha $\alpha_C=0.82$). Students' agreement ranging from 0 (low agreement) to 1 (high agreement) is visualised for all eight items ($^{\star}$items are inverted for scale analysis; error bars represent 1 SEM; items adapted from validated scales \cite{Altherr2003,Brooke1996, Unver2017}).
    }
    \label{fig:UD}
\end{figure}

\begin{figure}[t!]
\centering
\includegraphics[width=0.95\linewidth]{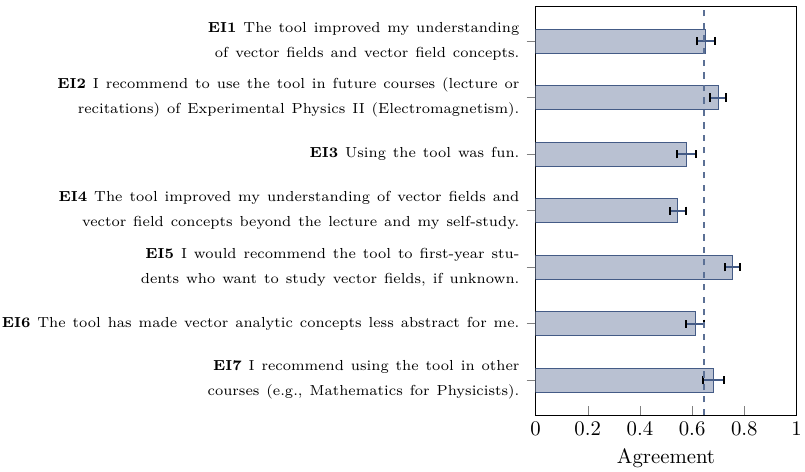}
\caption{Students' perceived educational impact of the graphical tool ($N=65$; Cronbach's alpha $\alpha_C=0.90$). Students' agreement ranging from 0 (low agreement) to 1 (high agreement) is visualised for all seven items (the dashed line indicates the scale mean; error bars represent 1 SEM; items adapted from a prior study \cite{Shellman2006}).
    }
    \label{fig:EI}
\end{figure}

The graphical tool enables to explore the important concepts of Gauss' and Stokes' theorems by one's own. However, providing an appropriate educational embedding with explicit tasks, that call for user actions, help to avoid misinterpretations and difficulties as well as support scientific working with the tool \cite{scientific}. For this purpose, we developed research-informed learning tasks, which are based on the visual interpretation of the vector field concepts proposed in Sects. \ref{sec.strategies} and \ref{Sec.Stokes}, integrate drawing activities, and involve the graphical tool in a structured manner (see \cite{Hahn2022, Hahn2023b} for conceptualisation of the learning tasks). The materials cover four units on divergence, Gauss' theorem, curl, and Stokes' theorem. Following common educational approaches, links between the units are established and the tasks are designed in parallel for both integral theorems \cite{Huang2013}. Hence, the units on divergence and curl refer to the left-hand side of Eqs. \ref{eq:Gauss} and \ref{eq:Stokes} and the units on Gauss’ and Stokes’ theorems correspond to the right-hand side of the equations and the connection between both sides. First, the tasks aim at formulating observations and drawing conclusions from the tool, for example, to discover divergence and curl as local properties of the field:
\begin{quote}
    \textit{Consider the vector field $\vec{C}(x,y)$ with \begin{equation*}
        \vec{C}(x,y) =xy\hat{e}_x+\hat{e}_y \end{equation*} \vspace{-6mm}\\ in the tool. For each of the four quadrants, evaluate how the field components of the vectors change along $x$ and $y$ directions, respectively, and conclude whether the divergence within each quadrant is positive, negative, or zero.} [learning task on divergence]
\end{quote}
Second, some tasks target at drawing vector field diagrams from the graphical tool to focus on characteristic properties of the representation \cite{Kohnle2020} and to examine conceptual relationships exemplarily from it. Moreover, some tasks ask for checking the results of a qualitative evaluation (e.g., of divergence) with the tool to connect different concepts and representational forms (diagram, formula).\\ 
In summer semester 2022 and 2023, we implemented the aforementioned multi-representational learning tasks in recitations of a weekly second-semester course on electromagnetism at the University of Goettingen and examined their impact on different outcome variables, for example, conceptual understanding, cognitive load, and representational competencies \cite{Hahn2023b}. Using a rotational design, students were first working with the multi-representational tasks and the tool and completed traditional, calculation-based tasks afterwards or vice versa (see \cite{Hahn2023b} for a detailed description of the study design and the methods). However, learning efficiency is not supposed to be the topic of this article, but we reveal first insights into students' usability and perceived educational impact of the tool. By taking together both studies, at the pretest, $N=199$ physics students (50 female) with a mean age of $20.3\pm2.1$ years (most of them in the second semester of study) participated in the studies. In summer semester 2022, they were learning with a desktop application (.exe) of the tool and in summer semester 2023, a web application was used.\\
After both intervention phases, we administered an evaluation questionnaire including eight items on usability and design \cite{Altherr2003,Brooke1996, Unver2017} and seven items on students' perceived educational impact of the graphical tool \cite{Shellman2006} (both scales with agreement on a 6-point Likert-type rating scale from 0 (low agreement) to 1 (high agreement)). Evaluation of students' perceived usability and design of the tool revealed a scale mean of $0.76\pm0.16$ (including item inverting) and students indicated high values for items that refer to user-friendliness, presentation form, and comprehensibility of the tool and its features ($N=125$; Fig. \ref{fig:UD}, UD1, UD3, UD5, UD6, UD7). Moreover, low values for items UD2 and UD4 suggest a good alignment of the graphical tool, the learning tasks, and students' prior knowledge, thus, we recommend to integrate the graphical tool in physics preliminary or introductory university courses. Moreover, as a first step, learners should become familiar with the design of the tool and its features. This can be done in a self-discovering way by entering different vector fields, describing and discussing all features and how the tool responds to changes. From the experiences of our implementations, we recommend starting with a constant vector field (e.g., $\vec{F}(x,y)=\hat{e}_x$), then moving to fields with one or two non-constant components that depend on the coordinates $x$ and/or $y$. To provide a more structured familiarisation, a short explanatory video introducing all features can be used (see e.g., \cite{Video1}). In Fig. \ref{fig:UD}, one recognises that about one third of the students reported technical issues with installing or using the tool, which were mostly related to the use of the desktop application in summer semester 2022 (UD8) and revised afterwards. \\
Further, students indicated a scale mean of $0.65\pm0.22$ for the perceived educational impact of the tool ($N=65$; Fig. \ref{fig:EI}). More precisely, the graphical tool is considered helpful for understanding (EI1, EI4, EI6) and recommended for first-year students and for its use in university courses in undergraduate physics (EI2, EI5, EI7). Additional discussion about students' experiences with the tool can provide guidance for its further revision and improvement (see Sect. \ref{sect.conclusion}). 

\section{Conclusion and outlook}\label{sect.conclusion}

We have presented an interactive learning environment that provides a qualitative approach to Gauss' and Stokes' theorems in two dimensions. The graphical tool takes several research-based learning difficulties about vector fields, divergence, curl, surface and line integrals into account, hence it can be considered as a starting point for learning about the integral theorems in a purely mathematical context. In contrast to traditional textbook instructions, the interactive and dynamic features allow students to explore vector fields and differential vector operators and, beyond existing visualisation tools and simulations, addresses both sides of Gauss' and Stokes' theorems as well as their relation. Based on the evaluation results, as next steps, we will extend the tool by including different surface shapes, for example, circles or hexagons to show that a divergence-free field remains divergence-free regardless of the shape of the test surface. We also plan to include several updates to the tool to approach the physics contexts of the theorem, namely an extension to curvilinear coordinates (particularly polar coordinates) and to three dimensions (including cylindrical and spherical polar coordinates). 
After a non-physical start based on Cartesian coordinates, these additions can support an approximation to physics problems; for example, the electric field of a point charge can be better described with polar coordinates than with Cartesian coordinates. Typical vector fields that correspond to real physics contexts can be included as a default. In addition, we plan to implement the graphical tool in further introductory physics and mathematics courses also to test the educational effectiveness of such revisions.  \\
Please note that there are also some limitations of the tool. Since it focuses on the concepts of Gauss' and Stokes' theorems, some basic misunderstandings of vector field representations are not fully addressed; for example, vector addition or unit vectors \cite{Barniol2014}. To address these concepts, we recommend using a learning environment that focuses on vectors and fields in general rather than on the integral theorems. Besides these limitations, the work presented here can serve as a basis for physics education and education research to build upon for accessing further concepts of vector calculus that are important in physics.

\section*{Acknowledgements}
We acknowledge the work of Robert M. Hanson from St. Olaf College in Northfield, MN \cite{Hanson} and thank him for providing inspiration and support with the tool in its early stages (\url{https://github.com/BobHanson/SwingJS-Examples/tree/SwingJS}).\\
The work is supported by the Open Access Publication Funds of the Göttingen University.

\section*{Competing interest/conflict of interest and ethical policy statement}
The authors declare that the research was conducted in the absence of any commercial or financial relationships that could be construed as a potential conflict of interest. Ethical review and approval was not required for the study on human participants in accordance with the local legislation and institutional requirements. The participants provided their written informed consent to participate in this study.

\section*{ORCID iDs}
Larissa Hahn: https://orcid.org/0000-0002-5864-1594\\
Pascal Klein: https://orcid.org/0000-0003-3023-1478

\end{document}